\documentclass[twocolumn,aps]{revtex4}
\usepackage{graphicx}
\usepackage{amsbsy}

\begin{document}

\title{Quasi-1D atomic gases across wide and narrow confinement-induced-resonances}
\author{Xiaoling Cui}
\affiliation{Institute for Advanced Study, Tsinghua University, Beijing, 100084 \\
Department of Physics, The Ohio State University, Columbus, OH
43210}
\date{{\small \today}}
\begin{abstract}

We study quasi-one-dimensional atomic gases across wide and narrow
confinement-induced-resonances (CIR). We show from Virial expansion
that by tuning the magnetic field, the repulsive scattering branch
initially prepared at low fields can continuously go across CIR
without decay; instead, the decay occurs when approaching the
non-interacting limit. The interaction properties essentially rely
on the resonance width of CIR.
Universal thermodynamics holds for scattering branch right at wide
CIR, but is smeared out in narrow CIR due to strong
energy-dependence of coupling strength.
In wide and narrow CIR, the interaction energy of scattering branch
shows different types of strong asymmetry when approaching the decay
from opposite sides of magnetic field. Finally we discuss the
stability of repulsive branch for a repulsively interacting Fermi
gas in different trapped geometries at low temperatures.
\end{abstract}

\maketitle

\section{Introduction}


Quasi-one-dimensional(1D) atomic gases across scattering resonances
can be realized in laboratories by utilizing the
confinement-induced-resonance(CIR)\cite{Olshanii}. By initially
preparing the system at high or low magnetic field and sweeping the
field properly, the quasi-1D system can evolve on the attractive
branch with molecules\cite{Moritz05} or on the repulsive scattering
branch that is free of molecules\cite{Haller11, Haller09}.
Quasi-1D atomic gases have many fascinating properties that are very
different from those of 3D gases. For instance, in quasi-1D a
two-body bound state exists for an arbitrary s-wave scattering
length $a_s$\cite{Olshanii, Moritz05}.
The three-body recombination rate of 1D bosons is efficiently
suppressed in the Tonks-Girardeau(TG) regime with strong
repulsion\cite{Gora03, Haller11}. A long-lived metastable quantum
phase in the super-TG regime\cite{Giorgini05,BA} has been realized
in the scattering branch of a 1D bosonic system with strong
attraction\cite{Haller09}.

Apart from dimensionality, the resonance width is another important
ingredient affecting many-body properties. Take (3D) Feshbach
resonance(FR) for example. In wide FR, where the width is much
larger than the typical energy scale $\mathcal{E}^*$\cite{E_scale},
the system exhibits universal thermodynamics(UT) right at resonance
where $a_s$ diverges\cite{Ho1,Thomas}. UT means that the
thermodynamic potential is a universal function of the temperature
and density, regardless of any detail of inter-particle
interactions. However, in narrow FR, where the width is much smaller
than $\mathcal{E}^*$, the universality is not evident due to
considerable effective-range effect\cite{Petrov-Lasinio-Peng}.
Another interesting property in narrow FR is that the interaction
energy shows strong asymmetry when approaching resonance from
different sides\cite{Ho3}, as recently been observed in $^6$Li Fermi
gas\cite{Ohara}. Considering the facts that the quasi-1D geometry is
reduced from 3D by confinement and the 1D resonance originates from
the 3D s-wave interaction, it is natural to expect the
effective-range effect in 3D will also influence the interaction
properties of quasi-1D system.

In this work, from the analysis of two-body solutions and
high-temperature Virial expansions, we study the scattering property
and thermodynamics of quasi-1D atomic gases across CIR. We shall
show that the quasi-1D geometry greatly modifies the stability of
repulsive scattering branch compared to 3D case. Similar to FR, CIR
can also be classified as wide or narrow according to the resonance
width. 
We find drastically different thermodynamic properties between wide
and narrow CIR due to the energy dependence of coupling strength.
The stability of repulsive branch in other trapped geometries, such
as isotropic 3D trap or anisotropic quasi-low-dimensional trapped
system, is also discussed in combination with recent developments on
cold Fermi gases in the laboratories. We use $\hbar=k_B=1$
throughout the paper.

The paper is organized as follows. In Sec. II, we introduce
effective quasi-1D scattering, from which the wide and narrow CIR
are defined and the corresponding thermodynamics are presented. In
Sec. III we carry out high-T Virial expansion for effectively 1D
system, and present a detailed study on the stability and
thermodynamic properties of the
repulsive branch. 
An extensive discussion of the stability of repulsive branch for a
cold Fermi gas in various trapped geometries is given in Sec. IV.
Finally we summarize the paper in Sec. V.

\section{Effective one-dimensional scattering and thermodynamics}

The Schr{\" o}dinger equation for the relative motion of two atoms
moving in quasi-1D is
\begin{equation}
H_0\psi(\mathbf{r})+ \frac{4\pi
a_s(E)}{m}\delta(\mathbf{r})\frac{\partial}{\partial
r}[r\psi(\mathbf{r})]|_{r\rightarrow0}
=E\psi(\mathbf{r}),\label{Hamil}
\end{equation}
here the non-interacting part is
$H_0=-\nabla^2_{\mathbf{r}}/m+m\omega_{\perp}^2(x^2+y^2)/4$,
$\mathbf{r}=(x,y,z)$ and $r=|\mathbf{r}|$; In the pseudopotential
part, we use the energy-dependent s-wave scattering length
obtained from a renormalization procedure\cite{KK,Yurovsky}.
\begin{eqnarray}
a_s(E)&=&a_{bg}(1+\frac{W}{E/\delta\mu-(B-B_0)}). \label{as}
\end{eqnarray}
$a_s(E)$ physically describe both wide and narrow FR, with
background scattering length $a_{bg}$, magnetic field $B$, resonance
position $B_0$, 
width $W$, and magnetic moment difference $\delta\mu$ between the
atom and closed molecular state.

The reduced quasi-1D scattering from 3D s-wave interaction has been
solved by Olshanii {et al}\cite{Olshanii}. For low-energy scattering
with $E=\omega_{\perp}+k^2/m$ and $k^2/m\ll 2\omega_{\perp}$, the
wavefunction at large inter-particle distance
is frozen at the lowest transverse mode, and its even-parity part is phase-shifted as
is $\Psi_{\rm even}(\mathbf{r})\sim\exp[-(x^2+y^2)/(2a_{\perp}^2)]\cos(k|z|+\delta_k)$, with $a_{\perp}=\sqrt{2/(m\omega_{\perp})}$ and
\begin{equation}
\cot\delta_k=-\frac{ka_{\perp}}{2}[\frac{a_{\perp}}{a_s(E)}-C_0+o(\frac{k^2a_{\perp}^2}{4})],
\label{1D_phase}
\end{equation}
where $C_0=1.4603$. Hereafter
we neglect the small correction from last term in Eq.\ref{1D_phase}.
$\delta_k$ in turn determines an 1D energy-dependent coupling
strength, $g(\bar{E})=2k\tan\delta_k/m$ with $\bar{E}=k^2/m$, as
\begin{equation}
g(\bar{E})=g_{bg}(1+\frac{W_{1D}}{\bar{E}/\delta\mu-(B-B_{1D})}), \label{g_E}
\end{equation}
where $g_{bg}=2\gamma\omega_{\perp} a_{bg},\ W_{1D}=\gamma W,\
B_{1D}=B_0-(\gamma-1)W-\omega_{\perp}/\delta\mu$, with
$\gamma=(1-C_0a_{bg}/a_{\perp})^{-1}$(see also \cite{Yurovsky}).
Eq.\ref{g_E} explicitly shows all realistic parameters describing
CIR, namely, the background coupling $g_{bg}$, resonance position
$B_{1D}$ and width $W_{1D}$. Near CIR ($B\sim B_{1D}$) and for
$\bar{E}\ll \delta\mu W_{1D}$, one can construct an effective-range
model to formulate 1D interaction,
\begin{equation}
\frac{1}{g(\bar{E})}=\frac{1}{g_{1D}}- \frac{m}{2\gamma^2\omega_{\perp}} r_0\bar{E}, \label{eff}
\end{equation}
with $g_{1D}$ the zero-energy coupling strength and
$r_0=-1/(ma_{bg}\delta\mu W)$ the effective range characterizing
E-dependence in $a_s(E)$\cite{r0}. To this end,
Eqs.(\ref{g_E},\ref{eff}) show the reduced effective-range effect(or
E-dependence of coupling strength) from 3D to quasi-1D system.


In the tight transverse confinements and low atomic density($n$)
limit, $n a_{\perp}\ll 1$, we consider an effective 1D system with
interaction given by Eq.\ref{g_E}. Generally, the pressure takes the
form
\begin{equation}
P=\mu (2m\mu)^{\frac{1}{2}}
\mathcal{F}(\frac{T}{\mu},\{\frac{\delta\mu(B-B_{1D})}{\mu},
\frac{\delta\mu W_{1D}}{\mu}, \frac{E_{bg}}{\mu} \}),
\label{P_general}
\end{equation}
where $\mu$ is the chemical potential, $E_{bg}=mg_{bg}^2$, and
$\mathcal{F}$ is a dimensionless function.

For wide CIR, $\delta\mu W_{1D}(\gg 2\omega_{\perp})\gg n^2/m$, the
E-dependence in Eq.\ref{g_E} and Eq.\ref{eff} is negligible, and the
interaction parameters in $\{...\}$ of Eq.\ref{P_general} can be
replaced by a single $g_{1D}$. The pressure is then reduced to
\begin{equation}
P=\mu (2m\mu)^{\frac{1}{2}}
\mathcal{F}(\frac{T}{\mu},\frac{\mu}{mg_{1D}^2}). \label{P_wide}
\end{equation}
At wide CIR($g_{1D}=\infty$), $P$ is just a function of $T$ and
$\mu$ (or $T$ and $n$) indicating UT for the scattering
branch\cite{att}. Particularly at $T=0$, UT can be established by
noting that the bosons and spin-$1/2$ fermions with infinite
repulsion are fully fermionalized, with the energy identical to that
of an ideal single-species Fermi sea\cite{BA,Guan}. However, at
narrow CIR($B=B_{1D}$), Eq.\ref{P_general} still essentially relies
on other interaction parameters ($W_{1D}, g_{bg}$) and thus UT is
absent. More explicitly, UT can be identified by Virial expansions
at high temperatures.

\section{High-temperature Virial expansion}

At high temperatures, $n^2/m\ll T\ll 2\omega_{\perp}$, we carry out
virial expansions on the effectively 1D system\cite{higher}. The
pressure can be expanded using the small fugacity $z=e^{\mu/T}$ as
$P=\alpha\frac{T}{\lambda} \sum_{n\geq1} b_n z^n$, where
$\lambda=\sqrt{2\pi/(mT)}$ is the thermal wavelength, $\alpha$ is
$1$ for spinless boson and $2$ for equal mixture of spin-$1/2$
fermions. Compared with non-interacting case(with superscript
$"0"$), 
\begin{eqnarray}
P&=&P^{(0)}+\alpha\frac{T}{\lambda} \sum_{n\geq2} (b_n-b_n^{(0)})
z^n. \label{dP}
\end{eqnarray}
Here the difference, $b_n-b_n^{(0)}$, characterizes the interaction
effect to the n-body cluster, which is generally a function of
$\{\frac{\delta\mu(B-B_{1D})}{T}, \frac{\delta\mu W_{1D}}{T},
\frac{E_{bg}}{T}\}$. For wide CIR, $b_n-b_n^{(0)}$ only depends on
one single parameter $1/(\lambda mg_{1D})$, which is free of
parameter at $g_{1D}=\infty$ for any order of Virial expansion and
leads to UT according to Eq.\ref{dP}.
This also justifies us in examining UT within the second-order
Virial expansion. Consideration of higher-order expansions will not
change the conclusion, except for a negligible correction (of higher
order in $z$ or $n\lambda$) to the thermodynamic quantities.

Due to the interaction effect, the second Virial coefficient,
$\Delta b_2=(b_2-b_2^{(0)})/\sqrt{2}$, can be written as $\Delta
b_2=\sum_l [e^{-E_l/T}-e^{-E_l^{(0)}/T}]$ (here $l$ is the energy
level for relative motion of two atoms). Given $P(T,\mu)$ in
Eq.\ref{dP}, it is straightforward to obtain the density $n=\partial
P/\partial\mu$ and entropy density $s=\partial P/\partial T$, and
finally energy densities, $\mathcal{E}=\mu n+Ts-P$, for spinless
bosons(b) and spin-$1/2$ fermions(f) as
\begin{eqnarray}
\mathcal{E}^b &=&\frac{
nT}{2}\Big[1+\frac{n\lambda}{2^{3/2}}(-1+2\epsilon_{int})+o((n\lambda)^2)\Big],\label{boson} \\
\mathcal{E}^f &=&\frac{
nT}{2}\Big[1+\frac{n\lambda}{2^{5/2}}(1+2\epsilon_{int})+o((n\lambda)^2)\Big],
\label{fermion}
\end{eqnarray}
with dimensionless interaction energy
\begin{equation}
\epsilon_{int}=-\Delta b_2+2T\frac{\partial\Delta b_2}{\partial T}.
\end{equation}

In the following we derive $\Delta b_2$ in strictly 1D by
enumerating the energy levels of two interacting particles in a
tube$([-L/2, L/2])$. For simplicity, we first consider the
scattering branch without inclusion of any bound state. The
discretized wavevector($k>0$) is determined by boundary condition
\begin{equation}
k_lL/2+\delta_l=(l+1/2)\pi\ \ \ (l=0,1,...).
\end{equation}
By comparing to non-interacting $k_l^{(0)}$ where $\delta_l=0$, we
obtain
\begin{equation}
\Delta b_2=\sum_{l}[ \exp(- k_l^2/(mT))- \exp(-k^{(0) 2}_l/(mT))],
\end{equation}
which can be transformed to an integral as
$2/(mT)\int_0^{\infty} dk k\delta_k e^{-k^2/(mT)}$ and further to
\begin{eqnarray}
\Delta b_2^{sc}&=&-\frac{1}{2}+\frac{1}{\pi} \int_{0}^{\infty} dk
e^{-\frac{k^2}{mT}} \frac{d\delta_k}{dk}.\label{b2}
\end{eqnarray}
Note that to obtain Eq.\ref{b2} we extrapolate $\delta_{l=0}$ to
$\delta_{k=0}$ in the thermodynamic limit, and set
$\delta_{k=0}=-\pi/2$ considering $\delta_{l=0}<0$ as well as
Eq.\ref{1D_phase}. 
When considering a bound state (occupy $l=0$), the lowest available
$l$ for scattering state should be $l=1$. This implies
$\delta_{k=0}$ is up-shifted by $\pi$ as revealed by Levinson's
theorem. In this case,
\begin{eqnarray}
\Delta b_2^{bd}&=&e^{-|E_b|/T}-\frac{1}{2}+\frac{1}{\pi}
\int_{0}^{\infty} dk e^{-\frac{k^2}{mT}}
\frac{d\delta_k}{dk}.\label{b2_bound}
\end{eqnarray}
To this end $\Delta b_2$ is obtained for both repulsive scattering
branch (Eq.\ref{b2}) and attractive branch (Eq.\ref{b2_bound}).
Remarkably, comparing with 3D\cite{Ho1}, $\Delta b_2$ in 1D has an
additional term $(-1/2)$ resulted from zero-energy phase shift and
the unique scattering property of 1D system.
Eqs.(\ref{b2},\ref{b2_bound}) are consistent with results obtained
from Bethe-ansatz solution\cite{Servadio} and analyses of real-space
wave functions\cite{Gibson}.

For quasi-1D system, the Virial expansions are carried out by
setting $k^{\Lambda}=2/a_{\perp}$ as an upper limit of integrals in
Eqs.(\ref{b2},\ref{b2_bound}). In the rest of this section, we shall
mainly focus on the scattering branch(cf Eq.\ref{b2}) which might
exhibit UT as discussed above.

\begin{figure}[ht]
\includegraphics[height=7.5cm,width=9cm]{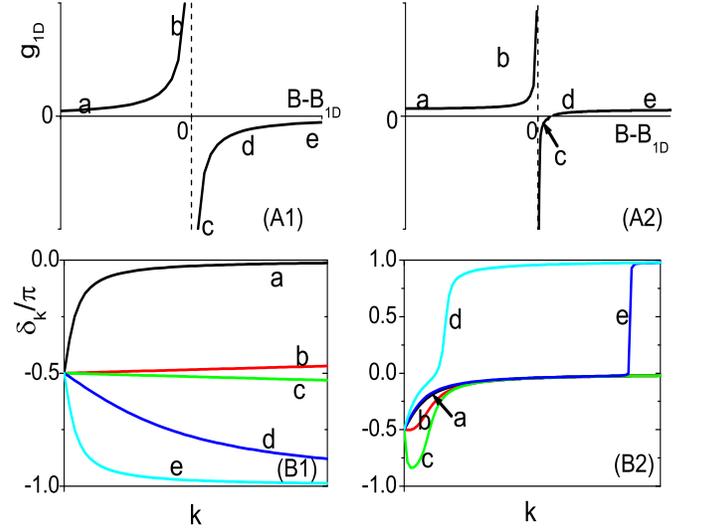}
\caption{(Color online). Upper panel: Schematic plot of zero-energy
coupling strength $g_{1D}$ across wide (A1) and narrow (A2) CIR. The
labels (a-e) correspond to $B\ll B_{1D}(a); B\rightarrow
B_{1D}-0^+(b); B\rightarrow B_{1D}+0^+(c); B> B_{1D}(d); B\gg
B_{1D}(e)$. Lower panel: Phase shift $\delta_k$ versus $k$ across
wide (B1) and narrow (B2) 1D resonances, with each label (a,b,c,d,e)
corresponding to specific $g_{1D}$ as marked in (A1) and (A2).}
\label{fig1}
\end{figure}


\subsection{Wide CIR}

With $\delta\mu W_{1D}(\gg 2\omega_{\perp})\gg n^2/m$, we replace
the E-dependent $g(\bar{E})$ by a constant $g_{1D}$.
$g_{1D}$ is schematically plotted in Fig.1(A1), giving the phase shift ($\delta_k$) of scattering
branch in Fig.1(B1). Here we have excluded the existence of
bound state for any B-field, thus $\delta_k$ all start from $-\pi/2$
at $k=0$. By increasing B across CIR(from "a" to "e"), the amplitude of $\delta_k$ at finite $k$ gradually becomes
enhanced, implying more repulsive energies in the system.
In particular, $\delta_k$ is uniformly $-\pi/2$ for all $k$ right at
CIR(between "b" and "c"), leading to universal values of $\Delta
b^{sc}_2$ and $\epsilon^{sc}_{int}$ as shown below.

\begin{figure}[ht]
\includegraphics[height=5.2cm,width=8.8cm]{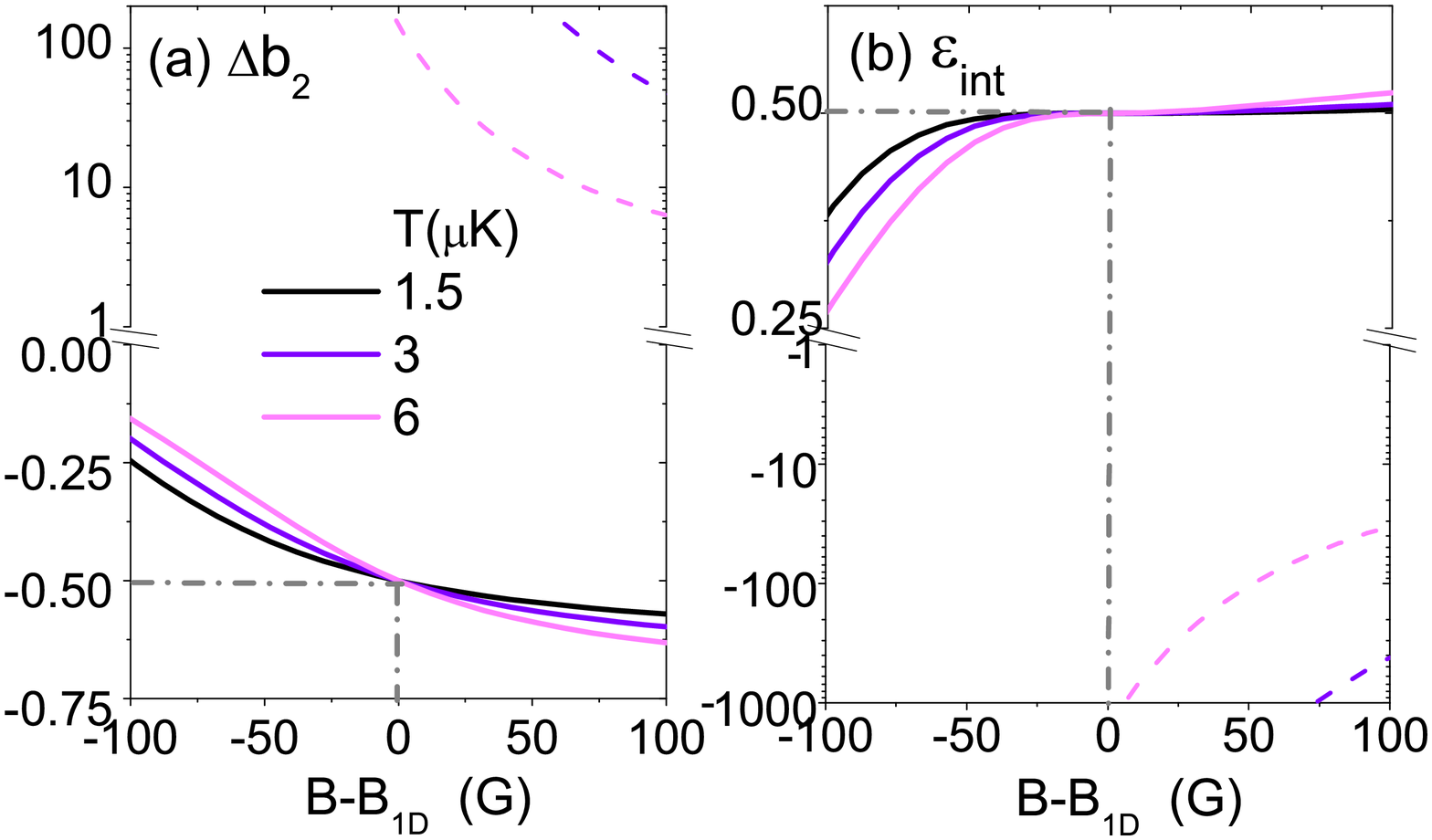}
\caption{(Color online). $\Delta b_2$(a) and $\epsilon_{int}$(b) for
two-species $^6$Li fermions across wide CIR at $T(\mu K)=1.5({\rm
dark\ black}), 3({\rm medium\ purple}), 6({\rm light\ pink})$. We
consider FR at $B_0=834.1G$ with width $W=-300G$\cite{Chin}.
Transverse confinements are generated by optical lattices with
lattice space $a_L=500nm$ and depth $V_0=25E_R$
($E_R=\frac{1}{2m}(\frac{\pi}{a_L})^2$), giving
$\omega_{\perp}=(2\pi)300$KHZ. CIR occurs at $B_{1D}=B_0-152.2G$
with $W_{1D}=-147.9G$, which satisfies $\delta\mu W_{1D}\gg
2\omega_{\perp}\gg T$. Solid and dashed lines are respectively for
scattering(Eq.\ref{b2}) and attractive(Eq.\ref{b2_bound}) branch.
Dash-dot lines denote universal values $-\Delta
b_2^{sc}=\epsilon_{int}^{sc}=1/2$ at CIR.}\label{fig3}
\end{figure}


For strictly 1D system with constant $g_{1D}$, Eqs.(\ref{b2},
\ref{b2_bound}) can be analytically solved, for example,
\begin{equation}
\Delta b_2^{sc}=-\frac{1}{2}+
\frac{sgn(g_{1D})}{2}\exp(\frac{1}{x^2})[1-erf(\frac{1}{x})],
\end{equation}
here $x=2\sqrt{2\pi}/(m|g_{1D}|\lambda)$; sgn() is sign function and
erf() is error function. In the weak coupling limit
($x\rightarrow\infty$), we obtain
\begin{equation}
\Delta b^{sc}_2=-1/(\sqrt{\pi}x),\
\epsilon^{sc}_{int}=2/(\sqrt{\pi}x)
\end{equation}
for scattering branch at $g_{1D}\rightarrow 0^+$ (corresponding to
solid lines in small B-field in Fig.2); and
\begin{equation}
\Delta b_2^{bd}=1/(\sqrt{\pi}x),\
\epsilon_{int}^{bd}=-2/(\sqrt{\pi}x)
\end{equation}
for attractive branch at $g_{1D}\rightarrow 0^-$\cite{m-f} (dashed
lines in large B-field in Fig.2). In the strong coupling limit
($x\rightarrow 0$), we obtain
\begin{eqnarray}
\Delta
b_2^{sc}&=&-\frac{1}{2}\pm\frac{1}{2\sqrt{\pi}}(x-\frac{x^3}{2}),\\
\epsilon_{int}^{sc}&=&\frac{1}{2}\mp\frac{x^3}{2\sqrt{\pi}};
\end{eqnarray}
the universal values at $x=0$, $-\Delta
b_2^{sc}=\epsilon_{int}^{sc}=1/2$, are direct consequences of
$k-$independent phase shift ($-\pi/2$) as mentioned above.

In Fig.2, we plot $\Delta b_2$ and $\epsilon_{int}$ for two-species
$^6$Li fermions across wide CIR. For scattering branch,
we see that all curves of $\Delta b_2^{sc}$(or
$\epsilon_{int}^{sc}$) at different $T$ intersect at a single point
in $\Delta b^{sc}_2$(or $\epsilon_{int}^{sc})$-B
plane\cite{correction}, demonstrating the UT of scattering branch
right at wide CIR.  The scattering system at strongly repulsive side
of CIR can smoothly evolve to strongly attractive side with even
higher energy. This is consistent with previous theoretical
predictions of sTG phase\cite{Giorgini05,BA} and its recent
experimental realization in bosonic gas\cite{Haller09}. Virial
expansion also shows that the scattering branch will achieve the
strongest repulsion as $g_{1D}\rightarrow 0^{-}$ at large B-field,
with $-\Delta b_2^{sc}, \epsilon_{int}^{sc}\rightarrow 1$. All above
properties can be clearly seen by tracing any individual energy
level of two scattering atoms in a tube, as shown in Fig.3(a).

Here we remark on the stability of scattering branch. In the
framework of two-body clusters in Virial expansion, the decay of
scattering branch manifest itself in the discontinuity of
thermodynamic quantities, due to the relabeling of scattering states
when the underlying bound state converts to the lowest scattering
state. This is why in 3D the decay occurs right at FR where the
bound state converts to scattering state at $a_s=\infty$\cite{Ho1}.
In 1D, however, the conversion is at $g_{1D}=0$ instead of at
resonance, and therefore the scattering branch can extend far away
from CIR until approaching zero coupling limit. A more comprehensive
discussion of the stability of scattering branch in other trapped
geometries will be given in Section IV.

\begin{figure}[ht]
\includegraphics[width=8.5cm]{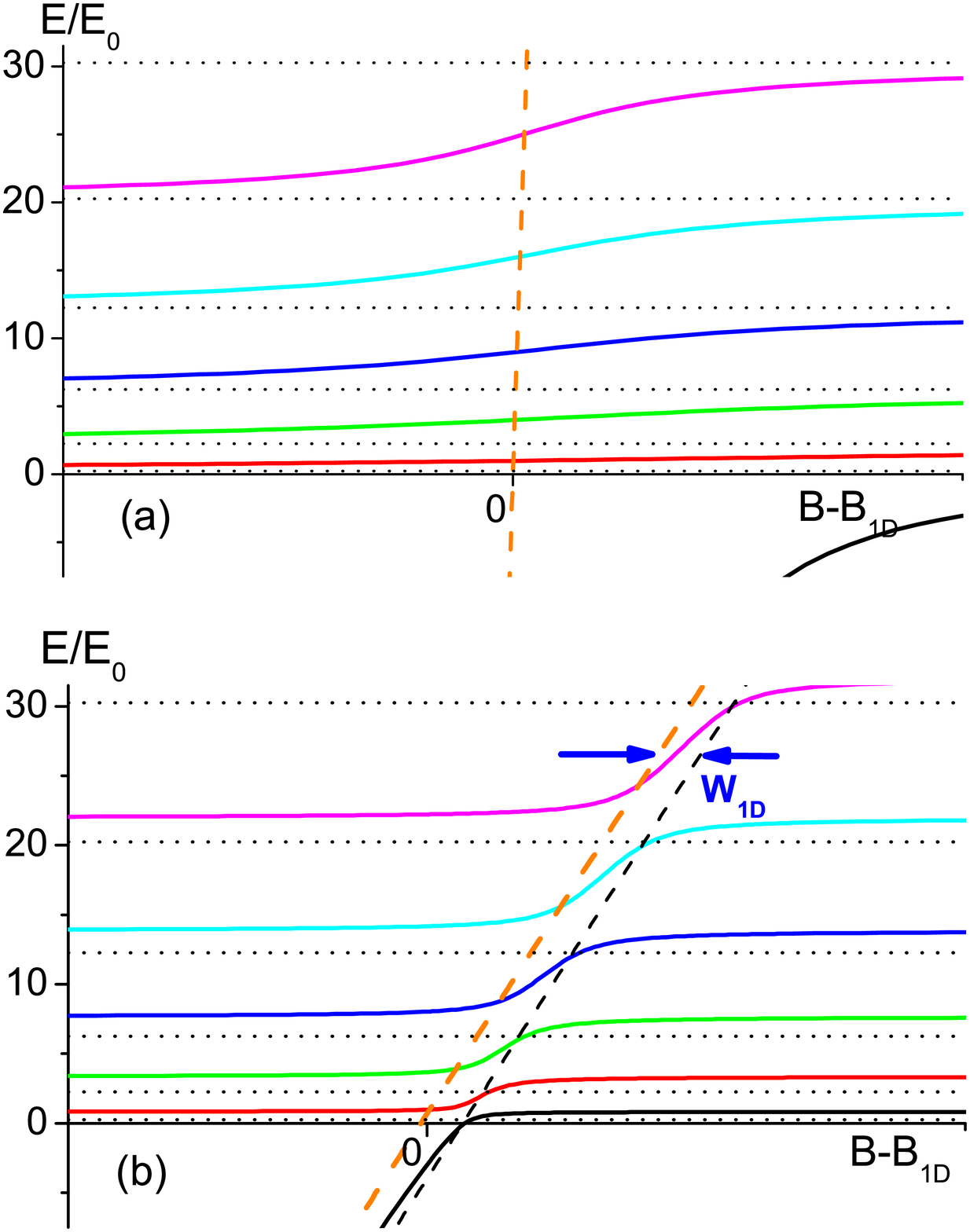}
\caption{(Color online). Two-body energy levels in the
center-of-mass frame for quasi-1D system confined in a
tube($[-L/2,L/2]$) across wide (a) and narrow (b) CIR. The orange
and black dashed lines denote $\pi/2$ and 0 phase shift
(corresponding to $g(E)=\infty$ and 0).
$E_0=(2\pi/L)^2/m$. The dotted lines
denotes non-interacting energy levels with
$E^{(0)}/E_0=(l+\frac{1}{2})^2,\ l=0,1,2...$.  }\label{fig3}
\end{figure}

At the end of this subsection we briefly discuss the second-order
virial expansion in a 1D harmonic trap, which can be carried out
given the two-body spectrum under coupling strength Eq.\ref{g_E}. In
particular, at wide CIR with $g_{1D}=+\infty$, the spectrum is
$E_l=(2l+3/2)\omega_z$ compared with $E_l^{(0)}=(2l+1/2)\omega_z$;
this gives $\Delta b_{2,trap}^{sc}=-1/(2\sqrt{2})$ compared with
$-1/2$ in homogenous case. In fact, based on local density
approximation(as used in 3D trapped system in Ref.\cite{Hu}), we
have
\begin{equation}
\Delta b_{n,trap}^{sc}=\frac{1}{\sqrt{n}}\Delta b_{n,hom}^{sc}
\end{equation}
for the scattering branch right at wide CIR. This shows a more rapid
convergence of virial expansions in trapped 1D system than in
homogeneous case.


\subsection{Narrow CIR}

With $\delta\mu W_{1D}\ll n^2/m (\ll \omega_{\perp})$, we take the
full form of $g(\bar{E})$ (Eq.\ref{g_E}) due to the strong
E-dependence. Assume a positive background $a_{bg}$, we give the
schematic plot of $g_{1D}$ in Fig.1(A2) and $\delta_k$ for
scattering branch in Fig.1(B2). In Fig.4, we show $\Delta b_2$ and
$\epsilon_{int}$ for $^{87}$Rb system across extremely narrow CIR.

Compared with wide CIR case, the scattering branch in narrow CIR
shows many distinct properties.

First, $\delta_k$ is no longer universal at CIR; instead, it
sensitively depends on $k$ and is quite small at finite-$k$ (see "b"
and "c" in Fig.1(B2)). This can be attributed to the strong
E-dependence in Eq.\ref{g_E} that $g(\bar{E})$ is far off resonance
at finite $\bar{E}$ even its zero-energy value $g_{1D}\rightarrow
\infty$. Accordingly, as plotted in Fig.3(b) the two-body levels at
finite energies are just shifted by a small amount although the
lowest level is shifted half-way. As a result, there is no UT at
narrow CIR, and the scattering branch is generally with very weak
repulsion even close to CIR (see also Fig.4).

Second, shortly beyond CIR, the scattering branch goes through a
decay at $B=B_{1D}+W_{1D}$ as manifested by discontinuous  $\Delta
b_2^{sc}$ and $\epsilon^{sc}_{int}$ there(see fig.4). This is
exactly the place where $g_{1D}$ evolves from $0^-$ to $0^+$ and the
bound state transforms to scattering state.

Third, after the decay, i.e., $B>B_{1D}+W_{1D}$  and $g_{1D}>0$, we
see from Fig. 1(B2) that $\delta_k$ will complete a continuous
change from $-\pi/2$ to nearly $\pi$ within an energy window $\Delta
\bar{E}\approx B-B_{1D}$. Due to the large $\pi$ shift for all
energies larger than $\Delta \bar{E}$, we see an large and negative
$\epsilon^{sc}_{int}$ in Fig. 4(b) despite of positive $g_{1D}$.
Similarly to narrow FR in 3D\cite{Ho3}, we expect the negative
$\epsilon^{sc}_{int}$ in extremely narrow CIR will extend to much
larger B-field, until $\delta \mu (B-B_{1D})$ approaches the typical
energy scale of the system\cite{E_scale}.

On the whole, the scattering branch in narrow CIR has weak
repulsion($\epsilon^{sc}_{int}\rightarrow 0^+$) or strong
attraction($\epsilon^{sc}_{int} \rightarrow (-1)^+$) when B
approaches the decay position ($B=B_{1D}+W_{1D}$) from the small or
large field side. The asymmetry here differs from that in wide CIR,
where $\epsilon^{sc}_{int}$ approaches $1^-$ or $0^+$ respectively.
It is also helpful to compare these features in quasi-1D with those
in 3D system\cite{Ho1,Ho3}. For wide FR in 3D, the amplitudes of
interaction energies are symmetric for the system evolving in
different branches and approaching FR from different
sides\cite{Ho1}, which is in contrast with what we find in quasi-1D
system across wide CIR. For narrow FR, the interaction effect are
greatly suppressed for repulsive branch but greatly enhanced for
attractive branch\cite{Ho3}, the same features as revealed above in
quasi-1D system across a narrow CIR. In cold atoms experiments, all
these features in quasi-1D system can be detected using the
technique of rf spectroscopy, as has been successfully applied to a
quasi-2D Fermi gas\cite{Kohl} and a 3D Fermi gas across narrow
FR\cite{Ohara}.

\begin{figure}[ht]
\includegraphics[height=5.5cm,width=9cm]{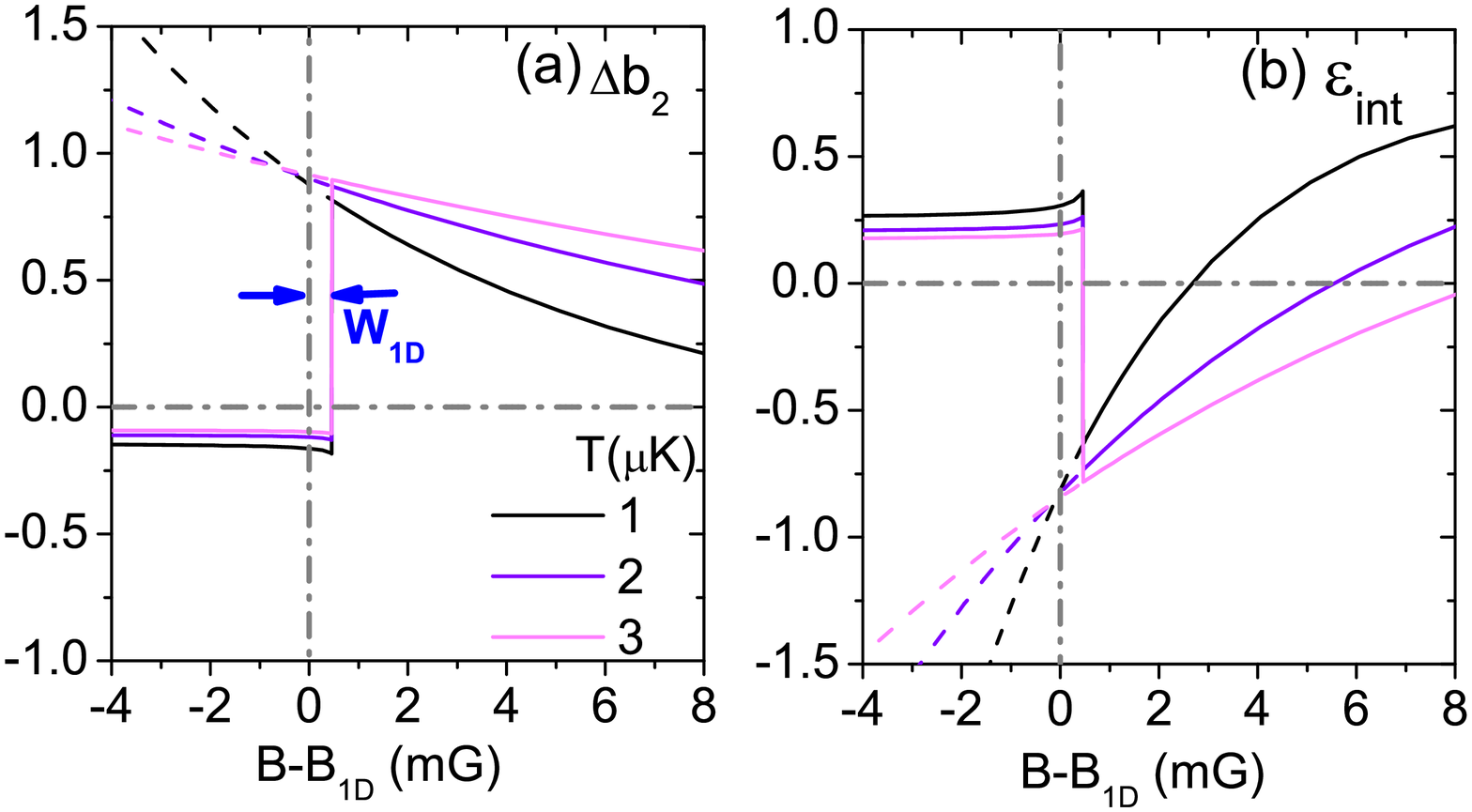}
\caption{(Color online). $\Delta b_2$(a) and $\epsilon_{int}$(b) for
$^{87}$Rb bosons across narrow CIR at $T(\mu K)=1({\rm dark\
black}), 2({\rm medium\ purple}), 3({\rm light\ pink})$. We consider
FR at $B_0=406.2G$ with width $W=0.4mG$\cite{Chin}. The optical
lattice is same as that for $^6$Li in Fig.2, giving
$\omega_{\perp}=(2\pi)80$KHZ. CIR occurs at $B_{1D}=B_0-27.1mG$ with
$W_{1D}=0.5mG$, which satisfies $\delta\mu W_{1D}\ll T\ll
2\omega_{\perp}$.
Decreasing $B$ across $B_{1D}+W_{1D}$, the scattering branch(solid
lines) continuously evolves to attractive branch(dashed lines) with
a bound state emerging at threshold.} \label{fig4}
\end{figure}

\section{Stability of repulsive Fermi gases in trapped geometries}

Recently, the metastable repulsive branch of atomic gases has
attracted lots of research interests, in the context of the
experiment by MIT group on itinerant ferromagnetism for a
repulsively interacting Fermi gas\cite{Ketterle1}. The same group
later claimed the absence of itinerant ferromagnetism from the
measurement of spin susceptibility, and attributed this to the
instability of repulsive branch against the molecule formation for a
3D Fermi gas near Feshbach resonance\cite{Ketterle2}. Similarly, the
instability of a repulsive Fermi gas has also been observed in the
quasi-2D Fermi gas\cite{Kohl}, but at negative $a_s$ side.

There have also been quite a few theoretical studies as to why the
repulsive Fermi gas in a 3D homogenous system is unstable close to
Feshbach resonance\cite{Combescot, Pekker, Vijay}.
For instance, the instability has been attributed to the shifted
resonance in the background of a Fermi sea\cite{Combescot}, the
pairing instability dominating over ferromagnetism
instability\cite{Pekker}, or the vanishing zero-momentum molecule
due to Pauli-blocking effect of Fermi-sea atoms\cite{Vijay}. All
these studies can lead to the same conclusion at low temperatures,
i.e., the 3D Fermi gas becomes unstable at the place where $a_s$ is
comparable to the inter-particle distance $(1/k_F)$, or
equivalently, the two-body binding energy $(E_b\sim1/ma_s^2)$ is
comparable to the Fermi energy $(E_F\sim k_F^2/m)$. The scattering
branch can only be stable when $E_b>E_F$, where the deep molecule
can not be absorbed by the Fermi sea atoms\cite{Combescot, Pekker,
Vijay}. In other words, in this parameter regime the existence of a
deep two-body bound state effectively stabilizes a many-body system
at the metastable repulsive branch. Since the physics behind this
criterion does not depend on any detail of the dimension or trapped
geometry, it should be equally applicable to other cases besides the
homogeneous 3D system. In the following, we will use this criterion
to study the stability of repulsive Fermi gas in various trapped
geometries at low temperatures.

Typically we consider three different types of trapping potentials,
namely, the isotropic or nearly isotropic 3D trap
($\omega_x\sim\omega_y\sim\omega_z$), the extremely anisotropic
quasi-2D ($\omega_z\gg\omega_x,\omega_y$) and quasi-1D trap
($\omega_z\ll \omega_x,\omega_y$). To facilitate the discussion, we
consider the system across wide resonance (with single interaction
parameter), while the extension to narrow resonance should be
straightforward.


In a trapped system, a two-body bound state is always supported no
matter how weak the attractive interaction
is\cite{Busch98,Olshanii,Petrov}. For a 3D isotropic trap
($\omega_x\sim\omega_y\sim\omega_z\sim\omega$), however, it should
be noted that the two-body binding energy $E_b$ at $a_s<0$ side is
less than the order of $\omega$, i.e., the level spacing of all
scattering states\cite{Busch98}. As a result, in the thermodynamic
limit with atom number $N\gg1$, $E_b(<\omega)$ at $a_s<0$ side is
negligible compared with the Fermi energy $E_F\sim N\omega$. The
system is then expected to behave similarly to the homogeneous case,
in a sense that the repulsive branch is stable only with positive
$a_s$ where the bound state is visibly deep ($E_b\sim N\omega$).
This is consistent with what have been observed in MIT
experiments\cite{Ketterle1,Ketterle2}.
On the contrary, for anisotropic quasi-2D or quasi-1D trap, the
energy spacing of scattering state is generally of the order of
trapping frequency of the shallow confinement, while the binding
energy can be of order of trapping frequency of the tight
confinement even at $a_s<0$ side\cite{Olshanii,Petrov}. For example,
for a quasi-2D trapped system at $a_s=\infty$,
$E_b\geq\omega_z\gg\omega_x,\omega_y$, and the existence of deep
molecule would be possible to stabilize the repulsive branch at
$a_s=\infty$ as long as $E_b>E_F\sim N\omega_{x,y}$. In this case,
the stable scattering branch can even extend to negative $a_s$ side,
as shown in the experiment with a  quasi-2D Fermi gas\cite{Kohl}.
The same conclusion can be drawn in the quasi-1D trapped case
($\omega_z\ll\omega_x,\omega_y$), which is also consistent with the
high-temperature result presented in the last section.

In short summary of this section, at low temperatures, the stability
of repulsive Fermi gas in a trapped geometry relies not only on the
existence of two-body bound state, but more importantly, on the
value of its binding energy compared with the typical energy scale
of a many-body system. In other word, here the two-body physics
should be evaluated in a many-body background. Generally, the
repulsive branch in quasi-low-dimensional systems is expected to be
more stable than that in an isotropic 3D system. Therefore the
low-dimensional system provides us a more favorable platform to
realize possible itinerant ferromagnetism in repulsively interacting
Fermi gases.

\section{Summary}

In this paper, we have studied the quasi-1D atomic gases across wide
and narrow CIR. Our main results are summarized as follows.

First, from high-temperature Virial expansions we obtain the
following:

(1) By tuning the magnetic field across CIR, the repulsive
scattering branch of quasi-1D system can evolve continuously across
CIR, from $g_{1D}=+\infty$ to $g_{1D}=-\infty$ side.

(2) Universal thermodynamics are identified for the repulsive
scattering branch right at wide CIR, but is found to be washed away
at narrow CIR by the strong energy-dependence of coupling strength.

(3) The decay of quasi-1D repulsive branch occurs when
$g_{1D}\rightarrow 0$. The interaction energy shows different types
of strong asymmetry between wide and narrow CIR, when approaching
the decay position from opposite sides of magnetic field.

Moreover, the second-order virial expansion presented in this paper
also serves as a benchmark for testing future experiment on 1D
atomic gases.

Second, we have discussed the stability of repulsive branch for a
repulsively interacting Fermi gas at low temperatures in different
trapped geometries. By evaluating the two-body bound state in the
presence of a Fermi sea, we conclude that the system can generally
be more stable in the quasi-low-dimensional trapped system than in a
3D isotropic trap. This should shed light on the current experiments
seeking for ferromagnetism in more stable and strongly interacting
Fermi gases in low dimensions.


The author is grateful to Tin-Lun Ho for stimulating discussions.
This work is supported in part by Tsinghua University Basic Research
Young Scholars Program and Initiative Scientific Research Program
and NSFC under Grant No. 11104158, and in part by NSF Grant
DMR-0907366 and by DARPA under the Army Research Office Grant Nos.
W911NF-07-1-0464, W911NF0710576.


\begin{thebibliography}{}
\bibitem{E_scale}The typical energy scale $\mathcal{E}^*$ is the temperature $T$ at high T
and the Fermi energy $E_F$ at low T.
\bibitem{r0} In the low-energy limit and near FR, one can expand $1/a_s(E)$ as $1/a_s(0)-mr_0E$. 
\bibitem{att}UT is absent for attractive branch at $g_{1D}=\infty$, where atoms tend to form deep molecules with typical size as small as $a_{\perp}$\cite{Olshanii,
Moritz05}. The corresponding energy scale is not within the scope of
this paper.
\bibitem{higher}All higher transverse modes ($n>0$) contribute a negligible
correction, which is of the order of $e^{-2\omega_{\perp}/T}$.
\bibitem{correction}The deviation, due to the finite cutoff ($k^{\Lambda}=2/a_{\perp}$) in quasi-1D system, is negligible if $T\ll 2\omega_{\perp}$.
\end{thebibliography}
\end{document}